\documentclass[runningheads]{llncs}

\usepackage[utf8]{inputenc}
\usepackage{pifont}
\usepackage{soul}
\usepackage[primitives,logic,advantage,operators,sets,keys,adversary,probability,notions,events,complexity,landau,asymptotics]{cryptocode}
\usepackage{xspace}
\usepackage{cleveref}
\usepackage{acronym}
\usepackage{colortbl}
\usepackage{booktabs}
\usepackage{hyperref}
\usepackage{wasysym}
\usepackage{pslatex}

\newcommand{\MA}{\mathcal{M}}
\newcommand{\adver}{\ensuremath{\MA}\xspace}
\newcommand{\A}{\ensuremath{\mathcal{A}}\xspace}
\newcommand{\B}{\ensuremath{\mathcal{B}}\xspace}

\newcommand{\G}{\ensuremath{\mathbb{G}}\xspace}

\renewcommand{\sk}{\ensuremath{\mathsf{sk}}\xspace}
\renewcommand{\pk}{\ensuremath{\mathsf{pk}}\xspace}
\newcommand{\Z}{\mathbb{Z}}
\newcommand{\kdf}[1]{\ensuremath{\mathsf{KDF}(#1)}}
\renewcommand{\mac}[1]{\ensuremath{\mathsf{MAC}(#1)}}

\newcommand{\sid}{\ensuremath{\mathsf{sid}}\xspace}
\newcommand{\txtmac}{\ensuremath{\text{MAC}}\xspace}

\newcommand{\pass}{\ensuremath{\pi}\xspace}
\newcommand{\tws}[1]{\mathsf{tws}(#1)}
\newcommand{\fpr}[1]{\mathsf{fpr}(#1)}

\newcommand{\cmark}{{\checkmark}}
\newcommand{\xmark}{{\ding{55}}}

\newcommand{\pep}{p$\equiv$p\xspace}

\acrodef{pakemail}[PakeMail]{PAKE-based Public Key Authentication and Key Exchange over Email}
\acrodef{uuid}[UUID]{universally unique identifier}
\acrodef{api}[API]{Application Programming Interface}
\acrodef{ea}[EA]{entity authentication}
\acrodef{km}[KM]{key management}
\acrodef{kms}[KMS]{key management system}
\acrodef{mitm}[MITM]{man-in-the-middle}
\acrodef{pep}[p$\equiv$p\xspace]{Pretty Easy Privacy}
\acrodef{otr}[OTR]{off-the-record messaging}
\acrodef{im}[IM]{instant messaging}
\acrodef{oob}[OOB]{out-of-band}
\acrodef{hisp}[HISP]{human interactive security protocol}
\acrodef{sas}[SAS]{short authentication string}
\acrodef{mac}[MAC]{message authentication code}
\acrodef{kdf}[KDF]{key derivation function}
\acrodef{dh}[DH]{Diffie-Hellman}
\acrodef{smp}[SMP]{socialist millionaires' problem}
\acrodef{ppss}[PPSS]{password-protected secret sharing}
\acrodef{pake}[PAKE]{password-authenticated key exchange}
\acrodef{mpc}[MPC]{multi-party computation}
\acrodef{pki}[PKI]{public key infrastructure}
\acrodef{ttp}[TTP]{trusted third party}
\acrodef{pq}[PQ]{Post-quantum}
\acrodef{ibe}[IBE]{identity-based encryption}
\acrodef{rom}[ROM]{random oracle model}
\acrodef{crs}[CRS]{common reference string}
\acrodef{icm}[ICM]{Ideal Cipher Model}
\acrodef{aam}[AAM]{Algebraic Adversary Model}
\acrodef{zk}[ZK]{zero-knowledge}
\acrodef{nizk}[NIZK]{non-interactive zero-knowledge}
\acrodef{dlp}[DLP]{Discrete Logarithm Problem}
\acrodef{dhp}[DHP]{Diffie-Hellman Problem}
\acrodef{cdh}[CDH]{Computational Diffie-Hellman}
\acrodef{csdh}[CSDH]{Computational Square Diffie-Hellman}
\acrodef{dsdh}[DSDH]{Decision Square Diffie-Hellman}
\acrodef{ddh}[DDH]{Decisional Diffie-Hellman}
\acrodef{didh}[DIDH]{Decision Inverted-additive Diffie-Hellman}
\acrodef{rlwe}[RLWE]{Ring Learning With Errors}
\acrodef{lwe}[LWE]{Learning With Errors}
\acrodef{svp}[SVP]{Shortest Vector Problem}
\acrodef{fs}[FS]{forward secrecy}
\acrodef{pfs}[PFS]{perfect forward secrecy}
\acrodef{kc}[KC]{key confirmation}
\acrodef{kem}[KEM]{Key Encapsulation Mechanism}

\definecolor{lightslategray}{rgb}{0.47, 0.53, 0.6}

\usepackage{fancyhdr}
\fancypagestyle{empty}
{
	\fancyhead{}
	 
	\fancyfoot{}
	\fancyfoot[L]{\scriptsize This is a copy of the author pre-print submitted to ICETE 2020-Extended and revised papers (Springer). }
	\fancyfoot[R]{\thepage}
}

\begin{document}

\title{PakeMail: authentication and key management in decentralized secure email and messaging via PAKE}
\titlerunning{PakeMail: authentication and key management in email and messaging via PAKE}

\author{Itzel Vazquez Sandoval\inst{1}, Arash Atashpendar\inst{1,2}, Gabriele Lenzini\inst{1}, Peter Y.A. Ryan\inst{1}}
\authorrunning{Vazquez Sandoval, Atashpendar, Lenzini \and  Ryan}

\institute{SnT, University of Luxembourg, Luxembourg\\
\email{\{itzel.vazquezsandoval,gabriele.lenzini,peter.ryan\}@uni.lu} \and
itrust consulting, Luxembourg\\
\email{atashpendar.arash@gmail.com}}

\maketitle

\begin{abstract}
We propose the use of \ac{pake} for achieving and enhancing \ac{ea} and \ac{km} in the context of decentralized end-to-end encrypted email and secure messaging, i.e., 
where neither a public key infrastructure nor trusted third parties are used.
This approach not only simplifies the \ac{ea} process by requiring users to share only a low-entropy secret, e.g., a memorable word, but it also allows us to establish a high-entropy secret key; this key enables a series of cryptographic enhancements and security properties, which are hard to achieve using \ac{oob} authentication.
We first study a few vulnerabilities in voice-based \ac{oob} authentication, in particular a combinatorial attack against lazy users, which we analyze in the context of a secure email solution.
We then propose tackling public key authentication by solving the problem of \emph{secure equality test} using \ac{pake}, and discuss various protocols and their properties. 
This method enables the automation of important \ac{km} tasks such as key renewal and future key pair authentications, reduces the impact of human errors, and lends itself to the asynchronous nature of email and modern messaging.
It also provides cryptographic enhancements including multi-device synchronization and secure secret storage/retrieval, and paves the path for forward secrecy, deniability and post-quantum security. We also discuss the use of auditable \ac{pake}s for mitigating a class of online guess and abort attacks in authentication protocols.
To demonstrate the feasibility of our proposal, we present PakeMail, an implementation of the core idea, and discuss some of its cryptographic details, implemented features and efficiency aspects.
We conclude with some design and security considerations, followed by future lines of work.

\keywords{Password-Authenticated Key Exchange \and Public Key Authentication \and Key Management \and Secure Email \and Secure Messaging \and Implementation \and Decentralized Trust Model}
\end{abstract}

\section{Introduction}
\label{sec:intro}
\noindent
Largely owing to cryptography, modern messaging tools (e.g., Signal) have reached a considerable degree of sophistication, balancing advanced security features, ranging from end-to-end encryption to forward secrecy and deniability, with high usability.
However, this has not been the case for email, even though it has a long history and remains the most pervasive and interoperable form of digital communication, with billions of emails exchanged on a daily basis \cite{clark2018securing}.
Yet, secure messaging and email share two long-standing challenges, namely entity authentication and key management.

Entity authentication, the primary concern, invariably involves a mechanism that associates some cryptographic material with an identity, e.g., public key authentication. Key management, affecting email more acutely, is intertwined with authentication and the need for automating it has been known for a long time, e.g., see \cite{ruoti2018comparative}.

Over the years, several methods have been established for accomplishing public key authentication, and indirectly key management: manual validation of key fingerprints, web of trust, \acf{pki} and hierarchical validation, public key directories as well as server-derived public keys such as \ac{ibe}. 

The set of viable techniques becomes much smaller once we consider a decentralized setting, i.e., without a \ac{pki} or a \acf{ttp}. 
In this context, approaches based on the use of \acf{oob} channels and \ac{sas} comparisons (see \Cref{sec:related-work}) have received a great deal of attention from the research community.
Due to the user interaction required in these approaches---e.g., manually verifying public key fingerprints---usability plays an essential role for a successful authentication. Therefore, reducing the gap between security and usability by finding optimal trade-offs has been a central theme of research for decades, with a plethora of long-standing open problems \cite{unger2015sokMessaging,clark2018securing}.

In an attempt to improve usability in the entity authentication process, Alexander and Goldberg \cite{alexander2007improved} proposed a modified solution to the \ac{smp} by Boudot et al. \cite{boudot2001fair}, also known as secure equality test, for authentication in the \ac{otr} protocol \cite{borisov2004off}.
To the best of our knowledge, this is the only work that proposes an approach for key validation, mainly suitable for online (synchronous) settings, that relies on users pre-sharing a low-entropy secret. 

Here we revisit the problem of public key authentication in a decentralized setting to propose a user-friendly and robust approach based on \acf{pake} for solving \ac{smp} via low-entropy secrets. These secrets are not expected to be sampled from a large, uniformly distributed space, but rather from a small set of values, e.g., typical human-memorable passwords or pin numbers. The task of \ac{smp} boils down to two parties verifying equality of their inputs $\pass_\A$ and $\pass_\B$ in a zero-knowledge manner, such that by the end, they learn nothing but the boolean result of the equality test.

Apart from offering improved usability properties and eliminating a host of vulnerabilities present in \ac{oob}-based protocols, as discussed in \Cref{sec:weakVoiceOOB}, we show how the \ac{pake}-generated secret key can be used to pave the path towards providing a series of cryptographic enhancements in secure email and messaging. 
These include automation in key management and key renewal, forward secrecy in a symmetric-key setting, deniability, post-quantum security, secure secret retrieval and storage, and auditability for mitigating a certain class of online guess and abort attacks in authentication protocols.

To demonstrate the feasibility of the proposed approach, we provide a complete implementation of the core ideas. This also shows how the suggested approach would not only work naturally in the context of secure messaging, but also in the inherently asynchronous setting of email.

By applying \ac{pake} to this problem, we advance the state-of-the-art in the use of shared low-entropy secrets for entity authentication, an idea considered only in \cite{alexander2007improved}. Moreover, while \ac{smp} is a subproblem solved naturally by \ac{pake}, the latter has not been applied to tackle the problem of authenticating public keys in decentralized settings. 

\subsection{Motivation}
Despite its crucial importance, secure email and messaging solutions tend to brush aside the entity authentication step. When provided, features allowing to authenticate contacts are not intuitively accessible\footnote{For example, in order to access the authentication menu in Signal or WhatsApp, users need to (1) select a chat (2) click on the contact's name (3) select ``View safety number/Encryption''.}, which contributes to users neglecting or being unaware of the process.
Solutions for the entity authentication problem in decentralized non-\ac{pki} environments, typically rely on users correctly executing a manual comparison, which has been repeatedly shown to be error-prone and inconvenient for users (e.g. \cite{kainda2009usability}).

Our incentive for replacing \ac{oob} authentication with a cryptographic protocol 
is to significantly reduce the impact of failures occurring in the authentication process: while failures in our cryptographic-based solution at most require users to repeat the process, failures in methods highly-dependent on user behavior could completely jeopardize the security of the communication.

Our motivation for using PAKE---a method that does not seem to have enjoyed enough recognition due to a lack of mature implementations, reluctance towards client side cryptography, patent-encumbered designs and perhaps even unawareness of its usefulness---is grounded not only in its independence from a \ac{pki} or a \ac{ttp}, but also in its provision of a \ac{zk} solution for the secure equality test problem using a low number of rounds, which makes it compatible with asynchronous settings, and in the fact that it enables additional cryptographic enhancements.

Additionally, the need for addressing common challenges such as key management automation and device synchronization spurred us on.
By implementing our \ac{pake}-based solution for entity authentication, we address two open problems in secure email and messaging \cite{unger2015sokMessaging,clark2018securing}: bridging the gap between known theoretical results and real-world solutions, and the need for more robust authentication methods that also improve the trade-off between security and usability in secure solutions.

\subsection{Contributions and structure}

A brief review of the state-of-the-art is covered in \Cref{sec:related-work}, followed by an overview of background concepts in \Cref{sec:background}.
In \Cref{sec:weakVoiceOOB}, we discuss a few vulnerabilities in the use of \ac{oob} channels for authentication, including a partial preimage attack targeting lazy users, which we analyze in the context of the \pep \cite{pepWeb} secure email solution.

Next, in \Cref{sec:pake} we describe how solving the \emph{secure equality test} using \ac{pake} leads not only to entity authentication but also to the establishment of a shared high-entropy secret key that can be used to achieve additional cryptographic tasks and properties.
We provide a concrete illustrative scheme along with an analysis of various \ac{pake} constructions and properties relevant for our work, and briefly analyze network transport mechanisms and security. 

In \Cref{sec:enhancements}, we elaborate on the said cryptographic enhancements, such as inattentive user resistance, automated key renewal, automated future key pair authentication and multi-device synchronization, along with security properties such as deniability, forward secrecy, post-quantum security, auditability for detecting guess and abort attacks, and secure secret storage and retrieval with applications in email and secure messaging.
\noindent
\paragraph{\textbf{Extended version.}} The main contributions of this work, which is an extended version of our previous paper \cite{secrypt20}, are as follows:
\begin{itemize}
    \item \ac{pakemail}: in \Cref{sec:pake-implementation}, we present a complete implementation of the main idea for a \ac{pake}-based authentication and key management approach in the context of decentralized secure email, serving as a proof of feasibility.
    The source code, along with the corresponding documentation, can be found at \cite{pakeMailGit}.
    \item In \Cref{sec:background,sec:pake} we provide more theoretical details on \ac{pake} protocols and cryptographic elements relevant for a concrete implementation.
    \item We provide an analysis in \Cref{sec:comparison}, comparing our proposal with state-of-the-art trust establishment approaches.
    \item We extend and improve our descriptions of the cryptographic enhancements in  \Cref{sec:enhancements}, including the notion of secure secret storage and retrieval,  a variant of which has been recently implemented for the Signal messaging system, but using a combination of different cryptographic constructions.
\end{itemize}
In \Cref{sec:security} we review the main security properties of our solution and elaborate on a few methods for low-entropy secret agreement and improving usability. We conclude in \Cref{sec:conclusions} with a more detailed outline of future directions and open questions.

\subsection{Related Work}\label{sec:related-work}
Unger et al. \cite{unger2015sokMessaging} and Clark et al. \cite{clark2018securing} provide extensive systematic surveys covering numerous aspects of secure messaging and email. We limit ourselves to the decentralized setting without elaborating on the drawbacks of web of trust approaches 
covered in the above mentioned works.

The literature contains a sizeable body of work on \ac{oob}-based approaches, considered first by Rivest \cite{rivest1984expose}, many of which are inspired by the original work of Vaudenay \cite{vaudenay2005secure} based on \ac{sas} comparisons, e.g., \cite{nguyen2011authentication,kainda2009usability,kainda2010secureMobile,tan2017can}, to name a few.
This area has also been investigated by the formal methods community, see e.g. \cite{delaune2017formal} for a recent formal analysis of \ac{sas}-based schemes in the symbolic model.

As for low-entropy secret-based authentication, to the best of our knowledge, the only work in the literature is by Alexander and Goldberg \cite{alexander2007improved} using a modified version of the SMP protocol by Boudot et al. \cite{boudot2001fair} for improving the \ac{oob}-based authentication process in \acf{otr} \cite{borisov2004off}. \ac{otr} is a cryptographic protocol originally designed by Borisov and Goldberg, aimed at enabling encrypted, authenticated and deniable instant messaging conversations with forward secrecy; this protocol was proposed as an alternative to PGP for ``casual'' conversations.

The variant proposed by Alexander and Goldberg sacrifices the fairness property of \cite{boudot2001fair} for efficiency and the authentication process requires \A and \B to be both online when entering their secret and for the subsequent exchange of messages.

\section{Framework and preliminaries}
\label{sec:background}
\noindent
We use \A and \B to refer to honest parties Alice and Bob, and \adver for the adversary, Mallory. We use $\sample$ to denote an element sampled uniformly at random, and $\concat$ to denote concatenation. We denote low-entropy secrets provided by users with \pass.

\textbf{Security model.}
We consider the standard Dolev-Yao model \cite{DolevYao}. We do not assume any additional trusted infrastructure. 
In one of our proposed methods for transport protocol, we assume the existence of untrusted buffer/relay servers, somewhat akin to the ones used in the design of Signal or OTR4 (see \Cref{sec:transport-mechanism}).
Regarding \ac{pake}s, we will consider several constructions in \Cref{sec:pake}, largely proven secure in the so-called BPR model \cite{bellare2000authenticated} under various hardness assumptions.  

\textbf{System requirements.}
Our proposal does not require any format modifications and preserves compatibility between existing email clients and servers; therefore, we assume standard requirements for email transfer.
As for secure messaging, we do not introduce any extra trust assumptions and no additional infrastructure requirements. Any exchanges relayed or buffered by intermediate servers can be done by untrusted ones.

\textbf{Cryptographic notions.}
Due to space limitations, we assume familiarity with common cryptographic concepts, in particular with  \ac{dh}-based computational hardness assumptions.

We discuss schemes based on the \ac{rlwe} problem, a special case of the \ac{lwe} problem whose security may be reducible to the hardness of solving the \ac{svp} in lattices, for which no efficient quantum algorithms are known, thus conjectured to be quantum-secure. \ac{pq} cryptography encompasses schemes that are considered to be safe against adversaries equipped with scalable, cryptographically relevant quantum computers.

We use $\kdf{s}$ to denote a key derivation function that takes a source $s$ of keying material, typically with a fair amount of entropy but not uniformly distributed, and produces one or more cryptographically strong secret keys, see \cite{krawczyk2010cryptographic} for details. We denote with $\mac{k,m}$ a keyed message authentication code scheme that computes a tag on $m$ under key $k$. We use ``Curve25519'' to refer to the underlying elliptic curve used in the elliptic-curve-Diffie-Hellman function by Bernstein \cite{bernstein2006}.

\textbf{Socialist Millionaires' Problem.}
In the realm of secure \ac{mpc}, Yao's millionaires' problem \cite{yao1982protocols} is a famous example in which two parties want to find out 
whose input is greater without revealing any more information on the actual value. \ac{smp} is a variant of this and a \ac{zk} proof of knowledge protocol, with the difference that the parties only wish to know if their inputs are equal.

A series of works has been dedicated to solving \ac{smp}, including a well-known solution by Boudot et al. \cite{boudot2001fair} that provides a fair and efficient protocol, where fairness roughly means that no party can evaluate the function and walk away with the result without the other party learning the output.

Garay et al. \cite{garay2004efficient} showed that the fairness and the security definition of \cite{boudot2001fair} are not compatible with the simulation paradigm and that their solution would not be secure when composed concurrently; they present a construction that can be composed arbitrarily, with similar complexity results.

\textbf{PAKE.}
Password-authenticated key exchange (\ac{pake}) protocols enable the establishment of secure channels without the need for a \ac{pki}, \ac{ttp} or empirical \ac{oob} channels. In essence, they address a secure two-party computation problem and allow two parties \A and \B who share only a low-entropy secret/password $\pass \in \mathcal{D}$, with $\mathcal{D}$ being some relatively small dictionary, to agree on a high-entropy cryptographic secret key $k$, 
using \pass for authentication. Since the seminal work of Bellovin and Merritt \cite{DBLP:conf/sp/BellovinM92}, numerous \ac{pake} protocols have been proposed, which largely fall into the two categories of balanced (symmetric) and augmented (or asymmetric), referred to as a\ac{pake}. The latter stores one-way mappings of passwords on the server side in client-server settings.

Intuitively, a core property of \ac{pake} is that a run of the protocol should not leak any information about the password.
Moreover, apart from protection against \ac{mitm} attacks and variants thereof such as replay/reuse and mixing attacks, they should also provide security against offline dictionary attacks by passive and active adversaries. While due to the use of low-entropy passwords, any \ac{pake} protocol is vulnerable to an online guessing attack, the goal is to ensure that at most one test per run constitutes the optimal attack strategy for an active \adver interacting with a party. 
Similar to \ac{smp}, \adver can mask failed guessing attempts as network failures, thus allowing numerous attempts without raising suspicion. 
This is in general unavoidable, however, we will see in \Cref{sec:pake} how a recent work by Roscoe and Ryan \cite{roscoe2017auditable} can mitigate this.

Most well-known \ac{pake} protocols rely on different variants of the \ac{dhp}, which means that their security is ultimately reduced to that of the \ac{dlp}. These typically make use of a cyclic group $\mathbb{G}$ of prime order $p$, generated by $g \in \mathbb{G}$, along with a hash function $H$, modelled as a random oracle, plus a few other public parameters, e.g., $M,N \in \mathbb{G}$ in the case of SPAKE2, as shown in \Cref{fig:pep-spake2}. Moreover, the passwords are viewed as elements of $\mathbb{Z}_p$ (obtained by hashing user passwords to some $\pi \in \mathbb{Z}_p$), which are used to blind the \ac{dh} terms by multiplying these terms by randomly chosen elements of $\mathbb{G}$ raised to \pass, e.g., $g^x \cdot M^\pass$, where $x \sample \mathbb{Z}_p$. The final session key is then derived by computing a hash of the entire transcript $\vec{T_P}$ of a run of protocol $P$, which includes all of the parties' public and private values, the user identities, the \ac{dh} terms and the password, i.e., $k = H(\pass, id_\A, id_\B, \vec{T_P})$.

Often, how passwords are agreed upon and the actual details pertaining to the exchange of user identities are left out, i.e., deferred to higher-level applications implementing the protocol.
It is typically required for a higher-level application to be able to refer to a session using a globally unique identifier, a channel binding often also called a session ID, which for technical reasons rooted in composability should be computed as a function of user instance roles and information exchanged over the network during the execution of the protocol, e.g., user IDs and public randomness. The session identification (ID) is usually defined as as the transcript $T_P$ of the communication/conversation between \A and \B, which can also be viewed as a random variable, being a function of the random values generated by \A and \B. These, among other things, protect against \ac{mitm}, unknown key share attacks and replay attacks.

\section{Pitfalls in out-of-band authentication}
\label{sec:weakVoiceOOB}
\noindent In \ac{oob} authentication, users typically compare some representation of a cryptographic
hash (fingerprint) of their partners' public keys via a separate authenticated channel. 
This representation is usually in the form of a list of words, numbers or images.

Strong security properties can be achieved if users execute the manual verification correctly.
Yet, the difficulty of having users do the corresponding tasks correctly while finding the right balance between usability and security is the root cause of security drawbacks, which have been amply discussed by research on fingerprint and \ac{sas} comparison via \ac{oob} channels (e.g., \cite{kainda2009usability,kainda2010secureMobile,tan2017can}).
Naturally, usability studies encourage the replacement of manual comparisons by automated software whenever possible \cite{tan2017can}. 
Some of the problems rooted in \ac{oob} authentication are as follows.

\textbf{Selection of an adequate \ac{oob} channel.}
In practice, the theoretical and strong authentication requirements of \ac{oob} methods are not easy to satisfy. While face-to-face conversations provide a strong authenticated channel \cite{nguyen2011authentication}, they are often not viable.
It is usually assumed that an \ac{oob} channel cannot be forged, but it can be blocked, overheard, delayed or replayed.
Typical instantiations are done via voice-based channels, e.g., a phone call.
However, some already consider voice-based \ac{sas} comparison to be obsolete from a security perspective \cite{unger2015sokMessaging} as nowadays messages can be forged by voice synthesizers with a small sample of the victim's voice.
Indeed, a voice impersonation attack on users comparing PGP words \cite{wiretapping2014} reported the fake voice to be indistinguishable in about 50\% of the cases.

\textbf{Social engineering attacks.}
Although there are multiple options for users to interact via \ac{oob}, little effort has been gone into designing precise protocols for humans to carry out the authentication process in a privacy-preserving and fair manner. This leads to various attack vectors based on misleading users as opposed to finding technical vulnerabilities.
For instance, without knowing the authentication value, \adver can fool \A into trusting her key by pretending to be \B, asking \A to read her fingerprint representation first, and then simply confirming that the fingerprints match.

\subsection{Inattentive users and partial preimage attacks}

\textbf{Inattentive and lazy users.} Here we consider users misreading words (inattentive) or comparing only subsets of them (lazy).
A recent paper by Naor et al. \cite{naor2018security}
analyzes approaches based on \ac{sas} authentication that are vulnerable to MITM attacks w.r.t. lazy users.
For instance, the approach in WhatsApp and Signal would be flawed if users compared only either the first or the second half of the value, since it would amount to verifying only one peer's fingerprint.
To fix this, the authors propose an influence spreading technique in which every bit of the value to be authenticated influences the generation of each element of the \ac{oob} representation.

\textbf{Partial preimage attack.}
Dechand et al. \cite{dechand2016empirical} study an attack aimed at finding a partial preimage for a fingerprint verified by lazy users; specifically, they assume that users check subsets of bits at the boundaries and in the middle.

We now give a more detailed description of their analysis. Let $p$ denote the probability of finding a partial preimage for a given fingerprint $f$ and $q$ its complementary event. 
To calculate $p=1-q$, we work out $q$ (i.e., the absence of partial preimages for a specific bit permutation).
Let $b$ be the length of the fingerprint $f$ and assuming that $r$ consecutive boundary bits are fixed (checked by the user), in this case, the leftmost and rightmost bits of $f$, we let $\ell$ denote the number of remaining bits in the middle from which a possible variation of $u$ bits could be fixed, i.e., checked by the user.

Thus, we have $2 \cdot r + u$ fixed bits that the adversary cannot invert without the user noticing. 
Valid preimages can thus be obtained by flipping up to $t = \ell - u$ bits within the middle bits; by removing these from the total space of size $2^b$, we obtain the number of invalid ones.
With $k$ denoting a given number of positions to modify, the valid strings are then given by $\binom{\ell}{k}$ choices of positions to flip. Thus, $q$ is given by
\begin{equation}
q = \frac{2^b-\sum_{k=1}^t\binom{\ell}{k}}{2^b}.
\end{equation}
Expressing $p$ as a function of the computational effort in terms of $e$ brute-force attempts, we have $p = 1 - q^e$. To estimate the number of steps needed for finding partial preimages with a success probability $\ge p$, we simply compute $e=\textrm{log}_q(1-p)$.
Expressing $e$ in base 2 gives results comparable to \cite{dechand2016empirical}.

\subsection{Case study}\label{sec:sec-of-pep-auth}
Pretty Easy Privacy (\pep) is a software aimed at providing usable privacy-by-default in email via end-to-end opportunistic encryption. 
The tool largely automates key management tasks.
The public key of a user is attached to outgoing emails when a key of the recipient has not been stored.
Received keys are automatically stored for future use (\emph{trust-on-first-use}) and  outgoing emails are automatically encrypted when a public key of the intended receiver is available.
This approach requires neither a PKI nor a TTP.

Similar to the PGP word list, \emph{\pep trustwords} \cite{pepTwds} are natural language words that two users compare via a low-bandwidth \ac{oob} authenticated channel to prevent \ac{mitm} attacks.
The trustwords generation algorithm $\tws{\cdot}$ is a deterministic algorithm that runs locally taking as input the public key of the peer obtained by email and the user's own public key. 
Informally, $\tws{\cdot}$ performs an $\XOR$ over the fingerprints of each of the input arguments, and then maps each block of 16 bits from the resulting 160-bit long string to a word in a predefined dictionary of size $2^{16}$, thus yielding a list of ten words. 

To encourage users to perform the \ac{oob} authentication, by default \pep shows only five words; this means that the peers compare the first 80 out of the 160 bits of a PGP fingerprint, assuming that they check all the words.
Since an ``influence spreading'' property, similar to Naor et al.'s, is already present, the best adversarial strategy is a brute-force attack over the public key space requiring $\bigO{2^{80}}$ steps to find a key $k$ such that the first 80 bits of $\fpr{k}$ are equal to those of $\fpr{\pk_B}$, with $\pk_B$ being the public key of \B.

We consider lazy users and compute estimates for partial preimage attacks similar to the one presented above.  
We consider the two cases where, out of five words, the user verifies $(i)$ the first and last words as well as two from the middle $(ii)$ the first and last words, along with one of the three in the middle. 
Let $b=80$, $\ell=48$ and for $(i)$ we have $u=32$ and we get $e \approx 2^{38}$; for $(ii)$, with $u=16$, we get $e \approx 2^{32}$. These results show that \adver would succeed with costs equal to and lower than the computational power estimated for an average adversary \cite{dechand2016empirical}.

Clearly the decision to show five words instead of ten by default needs to be reconsidered. Users might feel less annoyed by having to compare fewer words, however, its adverse effect on security is considerable as it practically renders brute-force attacks viable. 

\section{Authentication in secure email and messaging via PAKE}
\label{sec:pake}
\noindent 
We now show how \ac{pake} can be used to perform a secure equality test and thereby authentication, yielding a more efficient solution, compared to \ac{oob} methods and the \ac{otr} approach, with better security guarantees and further cryptographic features.

\textbf{Trust establishment using low-entropy secrets.}
For now, we assume that \A and \B share a low-entropy secret---e.g., a short password---either agreed upon beforehand or decided by posing and answering a question at the beginning of the mutual authentication.

Intuitively, the goal is for \A and \B to authenticate their public keys via a secure equality test of their respective secrets $\pi_A$ and $\pi_B$, without revealing any information about the latter; hence the need for a zero-knowledge protocol guaranteeing that upon termination of the protocol, the resulting transcript of the exchanges does not leak any information on $\pi_A$ and $\pi_B$, allowing \A and \B to learn only whether or not their respective secrets were equal.
In addition, it should not be possible for \adver to brute-force the password via offline dictionary attacks. Thus, \adver's only strategy would amount to making online attempts.

\subsection{Public Key Authentication via PAKE}\label{sec:pake-based-auth}

To determine at the end of a \ac{pake} run whether the user secrets $\pi_A$ and $\pi_B$ are equal, without revealing anything else,
we enforce explicit authentication using a \ac{kc} step after the key establishment phase. 
While this step may be optional in the general case for \ac{pake} protocols, here it would be necessary in order to bind the cryptographic material with an identity.
The information that \A and \B wish to authenticate---e.g., public key fingerprints for email addresses or phone numbers in Signal---can be incorporated either into the \ac{kc} phase or into the initial user secrets.

Next, we delve into the details of how this can be achieved using a concrete \ac{pake} protocol.
The literature contains several well-studied instances of \ac{pake}, therefore, we first pick a candidate to demonstrate how it can be used to validate public keys, and then compare a few prominent schemes according to specific properties of interest.
For the moment, we do not focus on engineering aspects related to (a)synchronicity and message transport mechanisms, but we will come back to these in \Cref{sec:transport-mechanism,sec:pake-implementation}.

\subsection{An Instantiation based on SPAKE2}\label{sec:instantiation-based-on-spake2}

\begin{figure*}[h]
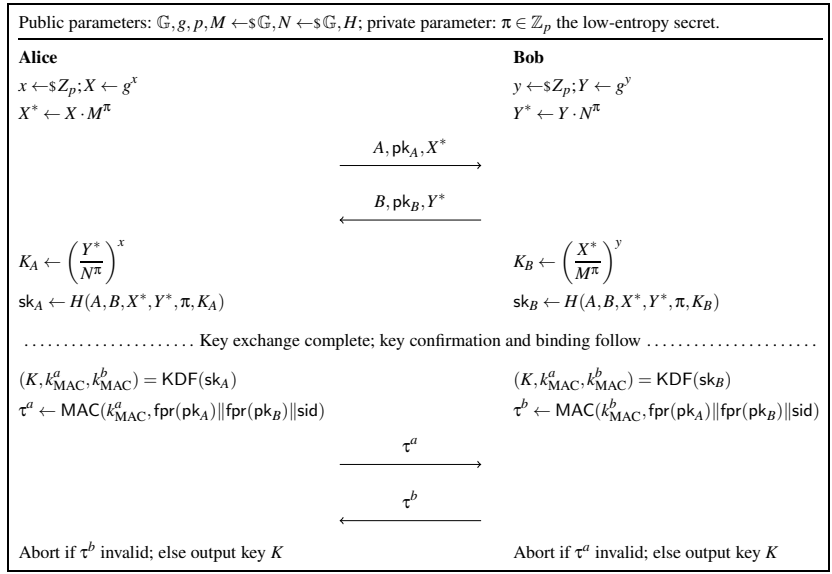

\centering
\resizebox{.9\textwidth}{!}{%
\fbox{
\procedure{Public parameters: $\G, g, p, M \sample \G, N \sample \G, H$; private parameter: $\pi \in \Z_p$ the low-entropy secret.}
{
\textbf{Alice} \> \> \textbf{Bob} \\
x \sample Z_p ; X \leftarrow g^x \> \> y \sample Z_p ; Y \leftarrow g^y\\
X^* \leftarrow X \cdot M^{\pi} \> \> Y^* \leftarrow Y \cdot N^{\pi} \\
\> \sendmessageright*[2.5cm]{A,\pk_A, X^*} \> \\
\> \sendmessageleft*[2.5cm]{B,\pk_B, Y^*} \> \\
K_A \leftarrow \left(\frac{Y^*}{N^{\pi}}\right)^x \> \> K_B \leftarrow \left(\frac{X^*}{M^{\pi}}\right)^y\\
\sk_A \leftarrow H(A,B,X^*,Y^*,\pi, K_A) \> \> \sk_B \leftarrow H(A,B,X^*,Y^*,\pi, K_B) \pclb
\pcintertext[dotted]{Key exchange complete; key confirmation and binding follow} 
(K,k^a_{\txtmac},k^b_{\txtmac}) = \kdf{\sk_A} \> \> (K,k^a_{\txtmac},k^b_{\txtmac}) = \kdf{\sk_B} \\
\tau^a \leftarrow \mac{k^a_{\txtmac},\fpr{\pk_A} \concat \fpr{\pk_B} \concat \sid} \> \> \tau^b \leftarrow \mac{k^b_{\txtmac},\fpr{\pk_A} \concat \fpr{\pk_B} \concat \sid} \\
\> \sendmessageright*[2.5cm]{\tau^a} \> \\
\> \sendmessageleft*[2.5cm]{\tau^b} \> \\
\text{Abort if } \tau^b \text{ invalid; else output key } K \> \> \text{Abort if } \tau^a \text{ invalid; else output key } K
}}
}
\caption{\pk authentication using SPAKE2 with refresh-then-MAC key confirmation for entity binding. (Originally presented as such in our previous work \cite{secrypt20})}
\label{fig:pep-spake2}
\end{figure*}

For illustration, in \Cref{fig:pep-spake2} we propose an extension of SPAKE2, a one-round protocol, with a \ac{kc} step to achieve explicit authentication, thus binding a public key to an entity. This yields a 2-round scheme, the minimum when \ac{kc} is enforced; see \cite{katz2011round} for optimal-round \ac{pake}s. For \ac{kc} we can use the generic refresh-then-\txtmac transformation. 
Despite its long history, this transform 
was only recently proved secure \cite{fischlin2016key}.

With $\G$ being a finite cyclic group of prime order $p$, generated by an element $g \in \mathbb{G}$, let $\G, g, p, M \sample \G, N \sample \G$ and hash function $H(\cdot)$ denote public parameters and $\pi \in \Z_p$ the private low-entropy secret, with the user password assumed to be appropriately mapped to an element in $\Z_p$. 
The parties perform the key exchange phase, as shown in \Cref{fig:pep-spake2},
which concludes with the generation of a symmetric key. Upon termination of the key establishment, \A and \B each use the symmetric key to carry out a key-refreshing step via a key derivation function in order to generate fresh \txtmac keys (for both parties), along with a new session key, $K$, which will be the final shared secret key. 
Next, under the freshly generated keys, they each compute a \txtmac on the fingerprints of both parties' public keys.
The authentication now amounts to exchanging and verifying the obtained tags $\tau^a$ and $\tau^b$, i.e., to see if the received tag and its locally computed counterpart match.

The addition of the \ac{kc} step increases the number of rounds and flows to 2 and 4, respectively. Note that this is merely an illustrative example and as already mentioned, other possibilities for \ac{kc} do exist, some of which offer additional properties. For instance, in \cite{becerra2018forward} the authors showed that a modified version of SPAKE2, called PFS-SPAKE2, coupled with a \ac{kc} step can achieve \ac{pfs} at the cost of increasing the number of rounds from 1 to 3. More recently, Abdalla et al. \cite{cryptoeprint:2019:1194abdalla} showed that SPAKE2 does indeed satisfy \ac{pfs} even without \ac{kc} under a different hardness assumption. They also prove a version with 
a \ac{kc} step (yielding a better bound) almost identical to the one given in \Cref{fig:pep-spake2}, except that the protocol has one less flow.

Alternatively, the public key fingerprints can be embedded in the secret $\pass$, but note that even in that case, the \ac{kc} step cannot be skipped as an explicit authentication of the public keys would be still needed. More precisely, we would let
$\pass = \pass' \concat \fpr{\pk_\A} \concat \fpr{\pk_\B}$,
where $\pass'$ denotes the original user provided secrets, and we would compute the tags as $\tau^a \leftarrow \mac{k^a_{\txtmac},\sid}$, where the session identifier \sid is defined as the transcript of conversation between \A and \B, with $\tau^b$ computed similarly.
The IETF documents for SPAKE2\footnote{\url{https://tools.ietf.org/id/draft-irtf-cfrg-spake2-08.html}} and J-PAKE\footnote{\url{https://tools.ietf.org/html/rfc8236}} provide similar one round \ac{kc} methods.

Note that the inclusion of $\pk_A$ and $\pk_B$ in the key exchange phase in \Cref{fig:pep-spake2} merely illustrates that they could be exchanged in one round. However, this exchange can be decoupled from the original SPAKE2 specification; indeed, the exchange of public keys may occur long before their authentication. This allows us to preserve the original description of the protocol and the computation of the transcript; otherwise, the key fingerprints would have to be included in the SPAKE2 transcript and in turn, in the input of the hash function computing the session key. In our case, the security guarantee is independent from this particular choice due to our strict enforcement of explicit authentication: fingerprints are included in the computation of the transcript (or session ID) in the \ac{kc} step.
We will elaborate further on this in \Cref{sec:pake-implementation}.

\subsection{Selecting a PAKE protocol}\label{sec:pake-choice}

We consider a number of representative \ac{pake} protocols and analyze their properties w.r.t. our use case: SPAKE2 \cite{abdalla2005simple}, OPAQUE \cite{jarecki2018opaque}, PFS-SPAKE2 \cite{becerra2018forward}, J-PAKE \cite{hao2010j}, KV-SPOKE \cite{katz2011round}, RLWE-PAK and PPK \cite{ding2017provably}. \ac{pake}s are typically evaluated according to the security model in which they are proven secure, support for forward secrecy, the number of rounds, along with their communication and computational complexity. The complexity related aspects become more relevant in a client-server setting wherein a server has to process a high number of requests and sessions in a short time span. In a decentralized peer-to-peer setting, such properties no longer play a major role.

In \Cref{tab:pake}, we present some relevant properties of the said constructions. Except for RLWE-PAK and RLWE-PPK that make use of lattice-based cryptography, all other schemes are Diffie-Hellman-based. In terms of \ac{pq} security, this implies that the latter cases would not be quantum-safe, whereas the first two would provide conjectured quantum-security due to the underlying \ac{rlwe} problem.

\begin{table}[t]
\centering
\footnotesize
\caption{Comparison of PAKE protocols. (Originally presented in \cite{secrypt20} )}
\label{tab:pake}
\begin{tabular}{|l|c|c|c|c|c|}
\hline
\centering Protocol & \multicolumn{1}{p{1.3cm}|}{\centering Rounds/ \\ Flows} 
             & KC
             & \multicolumn{1}{p{1.2cm}|}{\centering Forward \\ secrecy} 
             & \multicolumn{1}{p{1.7cm}|}{\centering Security \\ model}  
             & \multicolumn{1}{p{1.6cm}|}{\centering Hardness \\ assumption} \\ 
\hline
SPAKE2       & 1/2          & \xmark              & \cmark & ROM          & CDH    \\ 
PFS-SPAKE2   & 3/3          & \cmark              & \cmark & ROM          & CDH    \\
OPAQUE       & 2/3          & \cmark              & \cmark & ROM          & OMDH   \\
J-PAKE       & 2/4          & \xmark              & \cmark & ROM-AAM      & DSDH   \\
KV-SPOKE     & 1/2          & \xmark              & -      & CRS          & DDH    \\
RLWE-PAK     & 3/3          & \cmark              & \cmark & ROM          & RLWE     \\
RLWE-PPK     & 2/2          & \xmark              & \cmark & ROM          & RLWE   \\  
\hline
\end{tabular}

\smallskip
{\fontsize{8}{9} \selectfont
    ROM: Random Oracle Model; 
    AAM: Algebraic Adversary Model;
    CRS: Common Reference String
    
    DH: Diffie-Hellman;
    CDH: Computational DH; 
    DDH: Decisional DH;
    DSDH: Decision Square DH; 
    OMDH: One-More DH;
    RLWE: Ring Learning With Errors
}
\end{table}

Minimizing the number of rounds is more important for secure email than it is for messaging, especially if the transport mechanism is based on attachments or hidden emails (see \Cref{sec:transport-mechanism}).
As for secure messaging, this may be equally relevant for solutions that do not operate in a purely decentralized and peer-to-peer setting in which one may wish to reduce the load on relay or buffer servers, e.g., Signal or OTR4, but the number of rounds would in general be arguably less of a concern. Note that \ac{kc} can be added to schemes that do not have it by default at the cost of an extra round.

Intuitively, the notion of \ac{fs} captures the requirement that a long-term secret compromise should not result in prior session keys getting compromised and consequently the corresponding exchanges. Weak \ac{fs} (w\ac{fs}) refers to those schemes satisfying \ac{fs} against passive adversaries who did not interfere in the previous sessions and perfect \ac{fs} to those achieving the same against active adversaries. We will come back to this in \Cref{sec:crypto-properties}.

We limit the discussion on security models to practical considerations. In the \ac{rom}, an ideal truly random function being accessible to the parties through oracle calls is typically instantiated using cryptographic hash functions, and the \ac{crs} model implies the accessibility of a random string to all parties, generated in a trusted way. The latter may be less obvious to implement in the case of email due to the constraints of decentralization given that the generation of the \ac{crs} would be typically done by a trusted party or via a secure \ac{mpc} protocol, e.g., the decentralized \ac{crs} generation shown in \cite{sasson2014zerocash}.
Finally, regarding the RLWE-based schemes, their proofs are unfortunately in the \ac{rom}, as opposed to the quantum \ac{rom} (Q\ac{rom}), which would allow adversaries to query the random oracle in superposition.

\subsection{Transport Mechanism}\label{sec:transport-mechanism}

\paragraph{\textbf{Email-based approach.}} Given the small number of rounds required by PAKE protocols, in the case of email we can afford to use standard email attachments or specially formatted hidden emails as message carriers, processed in the background by the email client.
Since we primarily deal with authentication, these exchanges would have minimal impact in terms of communication and computational complexity as the protocol would have to take place only once per peer.

For the case of attachments, a \ac{pake}-based implementation could give \A the option to enter her secret $\pass_\A$ upon sending her first email to \B, thus allowing the first flow of the protocol to occur via an attachment; the initial \ac{pake} round would be completed when \B replies after entering his secret $\pass_\B$. The subsequent exchange for the \ac{kc} step can be done automatically.

Alternatively, the implementations could encapsulate cryptographic messages in specially crafted emails, kept hidden from the user (e.g., archived separately) and processed automatically---as \pep does for multi-device key synchronization.

\paragraph{\textbf{Untrusted server approach.}}
Although early \ac{im} tools were entirely online services that maintained an active session for each conversation, modern \ac{im} tools follow an asynchronous model similar to that of email.
For instance, both Signal and the latest version of \ac{otr} \cite{otrv4Spec} achieve offline messaging by using ``buffer servers'' for hosting pre-key bundles that can be fetched without the other party being online.

We can use a similar mechanism to overcome transport engineering obstacles in email more elegantly,
since all aspects related to the exchange of emails remain unchanged and thus interoperable. In fact, the use of an intermediate server would not introduce additional trust assumptions as the transcript of a PAKE protocol does not leak useful information to the adversary; such a server would be untrusted and any entity would be able to set up their own instance.

\section{Enhancements to secure email and messaging by \ac{pake}}
\label{sec:enhancements}
Our \ac{pake}-based approach for authentication satisfies and improves a number of key properties related to security and usability that have been identified in the literature \cite{unger2015sokMessaging}.
We first discuss how these properties are satisfied and then introduce novel uses of \ac{pake} in secure email and messaging. Note that once a \ac{pake}-generated symmetric key is established, subsequent \ac{pake} instances can be run automatically via a chaining self-sustaining mechanism; moreover, while we primarily focus on enhancements for existing paradigms that depend on public keys---e.g. PGP-based or OTR-inspired systems such as Signal---one could also consider the benefits of transitioning to symmetric-key constructions, e.g., \txtmac-based authentication and symmetric-key encryption schemes.

\subsection{Key management and authentication improvements}
The improvements presented here mainly deal with key management automation and error resilience.

\paragraph{\textbf{Automation of future key pair authentications.}}
Once authentication between \A and \B is bootstrapped from an initial \ac{pake}, the authentication of new key pairs belonging to either \A or \B can be automated by using the \ac{pake}-generated key as input, without prompting the users to yet again enter new secrets.
Authentication of new keys is needed for instance when keys expire, when a new key pair needs to be associated with an existing identity, or when new email addresses need to be associated with key pairs. These can be automatically authenticated by running a \ac{pake} with the stored shared symmetric key as input. Note that each execution of a \ac{pake} refreshes the stored \ac{pake}-generated symmetric keys.
Automating the authentication of future keys enables the achievement of the other properties in this category.

\paragraph{\textbf{Immediate enrolment.}}
This property holds if when a user reinitializes their keys, other parties can verify and use them immediately.
The \ac{pake}-generated key allows to automate the new key exchange and the corresponding authentication as explained above.

\paragraph{\textbf{Alert-less key renewal.}}
Complementing the previous property, this one refers to users not receiving alerts or warnings prompting them to take action when other parties renew their public keys. This would be automated similarly to immediate enrolment.

\paragraph{\textbf{Low key maintenance.}}
This property refers to minimizing users efforts related to key management tasks, such as signing keys or renewing expired keys. By achieving immediate enrolment and alert-less key renewal as explained above, the \ac{pake}-based approach improves key maintenance too.

\paragraph{\textbf{Inattentive user resistance.}}
As discussed earlier, manual \ac{oob} key/fingerprint verification methods are susceptible to human error and inattentiveness. In the \ac{pake}-based approach, even if users enter the wrong password, the result would not be as catastrophic as trusting a key prepared by the adversary. 
At worst, it would be inconvenient as the authentication would fail, prompting the user to eventually repeat the process.

\subsection{Cryptographic properties enabled by \ac{pake}}
\label{sec:crypto-properties}

\paragraph{\textbf{Symmetric key cryptography.}}
An immediate and rather evident advantage of using a \ac{pake} protocol in this context is that the resulting cryptographic secret key can be already used for performing cryptographic tasks, whereas in a general \ac{pki} setting, upon authenticating the public keys, one would then typically make use of a \ac{kem} in order to establish a symmetric key.

\paragraph{\textbf{Perfect forward secrecy (PFS).}}
Once, more popular in the context of secure messaging (e.g., Signal and OTR), \ac{pfs} is now a requirement for cipher suites supported in TLS 1.3. 
\ac{pfs} means that in the event of a password disclosure, previously derived session keys remain secure.
To minimize the impact of a long-term key disclosure, one could implement a \ac{pake}-chaining mechanism that automatically performs key rotations and periodically refreshes the symmetric key; this would provide limited windows of opportunity for \adver to compromise the channel, past which point, the fresh key would be secure again. If there is evidence that \adver has corrupted the channel, the cryptographic key would have to be discarded and replaced by a new \ac{pake} execution.
This refreshing paradigm might be expensive, however it would be relevant when \ac{pake}-based approaches are used for synchronization purposes, either device-to-device or device-to-server, where \ac{pake} can be used to both authenticate and establish a secure channel, thus providing \ac{pfs} for the session keys used for syncing. 

Several \ac{pake} constructions provide \ac{pfs} by default, some of which are listed in \Cref{tab:pake}; moreover, \ac{pfs} can be obtained by adding explicit authentication via a \ac{kc} step to constructions that do not have this property \cite{bellare2000authenticated}. 
Alternatively, to improve efficiency we could resort to symmetric-key schemes that provide \ac{pfs}, e.g., SAKE \cite{avoine2020symmetric}. In this case, a \ac{pake} can be used once to bootstrap authentication via a low-entropy secret and to generate the initial symmetric master key required by SAKE.

The use of \ac{pake}s could for instance improve the approach based on regular sub-key rotations, adopted by the Sequoia-PGP project for adding \ac{fs} to OpenPGP-based solutions; a \ac{pake}-based solution could automate authentication in case the master key, certifying the short-term sub-keys, needs to be refreshed.
For additional security, with slightly hampered usability, a separation of storage can be enforced by for example storing such \ac{pake} long-term keys in dedicated hardware, e.g., hardware security modules or smart key storage devices such as YubiKey or Nitrokey, to protect against a device compromise; see \Cref{sec:secret-retrieval} for more details on this.

\paragraph{\textbf{Deniability.}}
This is another subtle and fundamental property that has been of particular interest in recent secure messaging systems such as Signal and OTR. Deniable exchange, applied to tasks ranging from authentication to encryption, has a long and somewhat controversial history due to the subtleties in various existing security definitions. We limit ourselves to the case of key exchange and the seminal framework of Di Raimondo et al. \cite{di2006deniable}, which provides security definitions in the simulation paradigm for deniable key exchange and authentication, where both message and participation repudiation are considered as requirements. 

Assuming that the secret keys cannot be traced back to identities, we conjecture that sender/receiver-only deniability for symmetric \ac{pake} would satisfy the said definition of deniability in the symmetric-key setting:
in a two-party setup, a malicious accusing party \adver would not be able to produce binding cryptographic proofs from communication transcripts, associating another party with a particular exchange, as all messages could have been simulated by the accusing party \adver. More specifically, in terms of distribution indistinguishability, a simulator in the said framework \cite{di2006deniable} can be constructed given that $\pi$ is the only private input, symmetrically shared by both parties, and all other parameters are public and drawn at random. Indeed, this may not be surprising as Di Raimondo et al. \cite{di2006deniable} consider deniability in the symmetric key setting to be trivially satisfied.

Finally, assuming composability, using the \ac{pake}-generated key with symmetric ciphers and \txtmac-based authentication would preserve deniability. Clearly, this and other forms of deniability for \ac{pake} need to be studied rigorously in future work.

As a side note, deniability of messages and \ac{fs} were among \ac{otr}'s original goals, however, such features are independent from their \ac{smp} solution for authentication; they are implemented separately, e.g., by using \txtmac-based authentication and revealing keys. In the case of \ac{pake}, these properties are rather built into the scheme.

\paragraph{\textbf{Post-quantum security.}} 
As pointed out in \Cref{sec:pake-choice}, in the event that secure messaging and email tools transition to post-quantum cryptography, there are already candidate \ac{pake} constructions that provide conjectured \ac{pq} security (e.g. see \Cref{tab:pake}). 
Moreover, the recent symmetric-key authenticated key exchange (SAKE) by Avoine et al.\cite{avoine2020symmetric}  is conjectured to be \ac{pq}-secure due to its use of symmetric-key primitives. Thus, a quantum-resistant \ac{pake} can be combined with SAKE, to obtain a low cost and efficient \ac{pq}-AKE with \ac{pfs} suitable for settings with limited computational power, e.g., the IoT.

\subsection{Cryptographic enhancements to email and messaging}\label{sec:crypto-enhancements}
The uses of \ac{pake}s for securing email and messaging go beyond entity authentication and \ac{km}. Here, we discuss some areas that could benefit from the use of these schemes.

\paragraph{\textbf{Multi-device synchronization.}}
A quite natural application of \ac{pake} is in the realm of device pairing and secure multi-device synchronization, where the goal is to create an authenticated and private channel between devices, usually by the same user.
Most solutions typically rely on a \ac{hisp} and \ac{oob} channels, thus requiring manual intervention, which can give rise to new and subtle attacks. The application of \ac{pake}s for device pairing in other contexts has been considered before \cite{kumar2009comparative}; it is thus natural to consider its use in multi-device syncing of secure email and messaging systems, for instance, to synchronize a user's keys for encryption and keys of trusted contacts.

A secure email solution can display a screen in each of \A's devices that are to be paired, $D_1$ and $D_2$, so that after \A enters a password in both, a \ac{pake} protocol is triggered. Alternatively, this process can even be done asynchronously, i.e., without the two devices being online: $D_1$ pushes its state (e.g., key store, chat or email archive) to a server in encrypted form and later $D_2$ retrieves the secrets stored on the server in an oblivious manner w.r.t. the server. We discuss this further in the following part regarding secure secret retrieval.

For instance, the current implementation of \pep resorts to an ad-hoc pairing technique for key synchronization based on \ac{oob} comparison of \ac{sas}. Instead, it could benefit from such a \ac{pake}-based solution. The established channel could be used not only for sharing key material but also contact lists, calendars, etc.

\paragraph{\textbf{Secure Secret Sharing and Retrieval.}}
\label{sec:secret-retrieval}
This feature is inspired by the notion of \ac{ppss} schemes formalized by Bagherzandi et al. \cite{bagherzandi2011password}, which are $(t,n)$-threshold constructions wherein security is preserved against an adversary controlling up to $t$ servers out of $n$. A problem that \ac{ppss} addresses is protecting \A's secret data $d$ (e.g., a secret key used for decryption, authentication credentials, crypto-currency wallet key, etc.) in the event of a device compromise or failure.

An implementation of \ac{ppss} would secret-share $d$ among a set of $n$ entities so that only a collusion of more than $t$ corrupt ones would compromise the data. A password-based mechanism would allow the authentication of the owner of $d$ to the secret-share holders in order to trigger a reconstruction protocol and then retrieve the secret. 
The private storage of $d$ can be shared among $n$ external network entities; alternatively, if \A does not trust external entities, her device can instead partake in the secret-sharing by storing multiple shares, thus preventing online dictionary attacks by a network attacker and not allowing \adver to learn anything about the secret without corrupting \A's device.

Secret retrieval would have several use cases in secure messaging. For instance, a general anonymity/privacy related criticism directed at messaging services has to do with the identification of users via their phone numbers. This can be dealt with by securely storing long-term identities in encrypted form on the server, accessible only to the users.
Servers could also store per user lists of contacts in encrypted form; this would enable asynchronous syncing of contacts across multiple devices without the service provider learning the content.

Another use case would be to secret-share user data among several of their own devices, e.g., smartphone, laptop and tablet, so that a device compromise would not provide any useful information to an attacker; this can also be used for performing key synchronization among multiple devices. All these mechanisms would work in a similar manner from the user's point of view, i.e., simply by providing a password.

Recently, the Signal messaging system was enhanced with a functionality referred to as ``Secure Value Recovery'' \cite{signalRegistrationLock}, which aims at storing encrypted backups of user's data that can be recovered using a PIN. Among other things, the design involves a key stretching of the user’s PIN along with a master key derivation from the stretched key and a piece of server-side stored randomness. The same core functionality can be achieved with the use of either PPSS or PAKE constructions such as OPAQUE, a recent a\ac{pake} construction that, among other things, offers a secure secret retrieval mechanism based on oblivious pseudo-random functions, to fetch a secret stored in encrypted form on a server, using only a low-entropy password. It also offers protection against breaches and server password file compromises.

Signal’s developers also mention secret sharing and oblivious pseudo-random functions as future possibilities \cite{signalSecValueRecovery}, both of which could be achieved using existing cryptographic primitives, as explained above.

\paragraph{\textbf{Auditable \ac{pake}s for Thwarting Online Guessing Attacks.}}
\label{sec:auditable-pake}
As is the case for \ac{smp} in \ac{otr}, online guessing attacks are unavoidable in \ac{pake}s. This is usually dealt with by fixing a limit on the number of failed attempts that can be tolerated before invalidating a password.

However, in certain cases, another subtle adversarial strategy aimed at sidestepping the (at most) one online test per run would be to resort to a class of guess and abort attacks in which \adver intercepts a message in a given session (or initiates a session of her own) at a crucial step of a protocol run, verifies her guess at the password and in case of an incorrect guess, drops the said message to disguise her attempt as a network communication failure.

This can be done in both directions to double the chance of discovering the password, or in parallel against many network nodes depending on the setting. Such an attack can be carried out repeatedly without raising an alarm as the honest parties may simply view this as a network failure.

We identify a similar vulnerability in the use of a modified version of \ac{smp} in \ac{otr}: just before the last phase where the parties perform their secure equality test, when \A and \adver exchange their blinded \ac{dh} terms incorporating the low-entropy password in the exponent, i.e., $(g_3^a,g_1^a g_2^{\pass_\A})$, \adver could make a guessing attempt at $\pass_\A$ and in case of obtaining 0 (not equal), drop the message and force an abort, see sections 4.2 and 4.3 in \cite{alexander2007improved}. Note that the \ac{nizk} proofs that are attached to the messages at every exchange are not meant to protect against this type of attack.

In a relatively recent work, Roscoe and Ryan \cite{roscoe2017auditable} apply a mechanism based on commitment schemes and delay functions (e.g., timed-release encryption), originally developed by Roscoe \cite{roscoe2016detecting} for protecting against online attacks in \ac{hisp}s that use \ac{sas}, to the setting of \ac{pake}s in order to make them auditable by achieving \emph{stochastic fair exchange}.

Roughly speaking, this is achieved by a transformation for \ac{pake}s at the level of \ac{kc} using a combination of blinding, randomization, commitments and delay functions such that a series of messages consisting of fake ones and the real intended message are exchanged and the parties will only get to know which is the \emph{right} one until their exchange is complete. In a follow-up work, Couteau et al. \cite{cryptoeprint:2019:1281} generalize this result to achieve $\varepsilon$-fair exchange using oblivious transfer and timed-release encryption.

This transformation can be used to enhance any \ac{pake} with auditability, thus lending itself quite naturally to the authentication method suggested in this work.
An important limitation here is that, due to the highly interactive design of the solution, it would be more suitable to the setting of secure messaging than email, unless a given email solution were to opt for untrusted buffer servers for transport, see \Cref{sec:transport-mechanism}.

Finally, note that some of the ideas in this transformation, specifically those related to enforcing fairness, have common elements with the original \ac{smp} \cite{boudot2001fair} solution aimed at providing fairness, a property that was removed from the modified version of \ac{smp} used in \ac{otr} \cite{alexander2007improved} on account of achieving efficiency.

\subsection{Comparison}\label{sec:comparison}
\Cref{tab:features} shows a comparison of our proposal with a select set of approaches for trust establishment extracted from a relatively recent survey by Unger et al. on secure messaging \cite{unger2015sokMessaging}.
We limit our analysis to the most relevant aspects with respect to our proposal and refer the reader to the cited source for a more detailed explanation of the approaches and their properties.
If the reason behind a given evaluation is not specified in \cite{unger2015sokMessaging}, we provide our own interpretation and evaluate our approach accordingly.

\begin{table*}[t]
\newcommand*\rot[1]{\hbox to1em{\rotatebox[origin=bl]{65}{\textbf{#1}}} \hss}
\newcommand*\feature[1]{\ifcase#1 \xmark\or\LEFTcircle\or\CIRCLE\or\APLlog\or-\fi}
\newcommand*\f[3]{\feature#1&\feature#2&\feature#3}
\makeatletter
\newcommand*\ex[8]{#1&#2&%
    \f#3&\f#4&\f#5&\f#6&\f#7&\f#8&\expandafter\f\@firstofone
}
\makeatother
\newcolumntype{G}{c@{}c@{}c}
\caption{Comparison of trust establishment approaches. (Partial modifications to the original presented in \cite{secrypt20})}
\label{tab:features}
\centering
{\fontsize{8}{9} \selectfont
\begin{tabular}{lc G@{}G@{}G !{\kern0.5em} G@{}G@{}G !{\kern0.5em} G !{\kern1.5em}}
\toprule
Paradigm  & Example & \multicolumn{9}{c}{Security} & \multicolumn{9}{c}{Usability} & \multicolumn{3}{c}{Adoption}\\
\midrule
&& \rot{Network MitM Prevented}
 & \rot{Operator MitM Prevented}
 & \rot{Operator MitM Detected}
 & \rot{Operator Accountability}
 & \rot{Key Revocation Possible}
 & \rot{Privacy Preserving}
 & \rot{Deniability Facilitated}
 & \rot{Forward Secrecy Facilitated}
 & \rot{Post-quantum Security}
 & \rot{Automatic Key Initialization}
 & \rot{Low Key Maintenance}
 & \rot{Easy Key Discovery}
 & \rot{Easy Key Recovery}
 & \rot{In-Band}
 & \rot{No Shared Secrets}
 & \rot{Alert-less Key Renewal}
 & \rot{Immediate Enrollment}
 & \rot{Inattentive User Resistant}
 & \rot{No Service Provider}
 & \rot{Asynchronous}
 & \rot{Multiple Key Support}\\
\midrule
\ex{Web of Trust}{PGP}   {222}{110}{040} {001}{100}{000} {222}\\
\ex{KD + SaL}{CONIKS}    {201}{222}{040} {222}{222}{222} {222}\\
\ex{OE + TOFU}{TextSecure}  {111}{102}{444} {222}{222}{020} {220}\\
\ex{OE + TOFU + \ac{oob}}{\pep}  {111}{112}{444} {212}{102}{000} {200}\\
\ex{OE +  \ac{smp}}{OTR}  {111}{122}{440} {020}{011}{020} {202}\\
\ex{\cellcolor{lightslategray!65}OE + PAKE} {\cellcolor{lightslategray!65}PakeMail}  {113}{312}{333} {222}{220}{222} {222}\\
\ex{KFV: \ac{oob}}{SilentText}  {222}{212}{444} {000}{002}{000} {200}\\
\ex{KFV: \ac{smp}}{OTR}  {222}{212}{440} {000}{020}{000} {200}\\
\ex{\cellcolor{lightslategray!65}KFV: PAKE}{\cellcolor{lightslategray!65}PakeMail}  {222}{212}{223} {222}{220}{222} {222}\\
\bottomrule
\end{tabular}
}

\scriptsize
\smallskip
The property is: $\feature2=\text{satisfied}$;
     $\feature1=\text{partially satisfied}$;
     $\text{\feature0}=\text{not satisfied}$;
     $\feature3=\text{implementation dependent}$;
     $\text{\feature4}=\text{N/A}$
     
KD = Key directory; KFV = Key fingerprint verification; OE = Opportunistic encryption; SaL = Self-auditable logs; TOFU = Trust-on-first-use
\end{table*}

Most of the properties have self-explanatory names, except perhaps operator accountability, which is considered to be satisfied if the paradigm provides support for verifying the correct behavior of service providers during the trust establishment process, when a centralized infrastructure is required. The network and operator attackers considered for \ac{mitm} refer respectively to adversaries controlling large segments of the internet and infrastructure operators (service providers).

\ac{pake}-based approaches satisfy privacy preservation as the transcript of a \ac{pake} execution does not leak information. \emph{Deniability facilitated, FS facilitated} and post-quantum security are subject to the selection and exact usage of the PAKE scheme. 

Approaches built upon opportunistic encryption (OE) partially provide \ac{mitm} prevention because an attack can be successful during the initial communication round, before a key is authenticated. When combined with SMP,
operator accountability and \ac{mitm} detection are also partially satisfied given that if the execution of the SMP protocol fails, the users do not learn whether this was due to mismatching passwords or an adversarial attempt at compromising the channel. However, when it comes to our \ac{pake}-based approach, these last two properties could be potentially satisfied with the use of auditable PAKEs (see \cref{sec:auditable-pake}), mainly in the context of messaging.

It is somewhat ambiguous as to why the authors of \cite{unger2015sokMessaging} consider key revocation---users being able to revoke and renew keys---to be fully satisfied for SMP applied to OE. While revocation is possible, the process would still suffer from the known limitations of a truly decentralized setting, e.g., informing all users of an expired key. 
The latter is indeed stated to be the reason for considering that KFV approaches only partially satisfy this property.
Therefore, \ac{pake} applied to OE would also partially satisfy key revocation. Thanks to the derived cryptographic key, the main advantages of OE with PAKE can be observed at the level of usability related properties, e.g., automation of tasks.

In key fingerprint verification (KFV) approaches, the verification is considered to occur before using the public keys, which leads to achieving most of the security properties. The evaluations for the \ac{oob} approach assume that the manual comparison is executed correctly; this assumption is not needed for \ac{smp} or \ac{pake}.
As we can observe, \ac{pake}-based KFV significantly improves usability compared to \ac{oob} and SMP fingerprint verification. 

Key directory combined with self-auditable logs (KD+SaL) is arguably the most promising approach identified by Unger et al. due to the wide range of properties that it provides. It allows users to efficiently verify the consistency of their own entry in a central key directory and therefore to detect and expose misbehavior by a third party. 

The set of properties that KD+SaL and KFV:\ac{pake} can achieve is similar, yet, the latter has the advantage of enhancing security with the properties discussed in \Cref{sec:crypto-properties}. Overall, \ac{pake}-based key fingerprint verification offers the most complete set of properties with reasonable trade-offs between security and usability in a purely decentralized setting.

Clark et al. \cite{clark2018securing} present a similar table evaluating primitives used to enhance email security. Considering end-to-end encryption as a baseline, \ac{pake}-based key verification/management would perform as shared secret key verification (R14 in \cite{clark2018securing}), except that, additionally, our PAKE-based approach partially satisfies the property that refers to providing support for server-side content processing (P12) as this can be enabled without exposing the encrypted content, e.g., via secure secret retrieval (see \cref{sec:secret-retrieval}).

\section{Implementation: \ac{pakemail}}
\label{sec:pake-implementation}
Here we present \ac{pakemail}, an implementation of the core set of features of our proposal, mainly aimed at demonstrating the feasibility of the key ideas presented in this work. The source code and related documentation are available at \cite{pakeMailGit}.

\ac{pakemail} is a complete implementation of the main functionalities, namely, carrying out a \ac{pake} protocol in a decentralized setting to authenticate public keys and establish a shared symmetric cryptographic key, using standard email and attachments as transport mechanism for networking, while preserving interoperability and without introducing any extra trust assumptions. However, this implementation should be rather viewed as a proof of concept given that a full-fledged version would not only require additional design and security considerations, but it would also provide support for the other remaining features that we have discussed in \Cref{sec:enhancements}.

Our solution is implemented in Python 3, specifically targeted at version 3.6, with minimal dependencies, largely using standard Python libraries for tasks such as email formatting (MIME), encoding and exchange (IMAP, SMTP, TLS) as well as networking and file system operations. In terms of design, we have mainly adopted an object-oriented programming paradigm, enabling well-established properties such as a modular implementation with better separation of concerns via encapsulation, extensibility and re-usability. The current implementation is geared towards Unix-like operating systems, but it can be easily ported to other platforms.

\subsection{Cryptographic details}

\ac{pakemail} makes use of the SPAKE2 library developed by Warner \cite{spake2python}, which by default uses ``Curve25519''\footnote{\url{https://mailarchive.ietf.org/arch/msg/cfrg/-9LEdnzVrE5RORux3Oo_oDDRksU/}} for the underlying elliptic curve, offering 128 bits of security. It is however possible to switch to 1024/2048/3072-bit integer groups as well. 
For the key confirmation phase described in \Cref{fig:pep-spake2}, we use HKDF (HMAC-based Extract-and-Expand Key Derivation Function)\footnote{\url{https://tools.ietf.org/html/rfc5869.html}} by H. Krawczyk for implementing the key derivation function, and HMAC\footnote{\url{https://tools.ietf.org/html/rfc2104.html}} keyed-hashing for message authentication to derive the authentication tags.
Finally, we use the PyNaCl library, which is a wrapper for the well-known NaCl library, for performing cryptographic tasks such as encryption using 256-bit \ac{pake}-derived secret keys. 

\ac{pake} messages and passwords are stored and transferred as byte strings. While an encoding at the application layer can be applied, ultimately, the underlying SPAKE2 \ac{api} requires byte strings, thus leaving such choices to the users of the library. Moreover, due to the inherently asymmetric design of the SPAKE2 implementation, we assign distinct roles to \ac{pake} instances, which in our implementation are referred to as ``initiator'' and ``responder''. Also, among other things, to prevent message reuse in different contexts and in line with the original protocol description \cite{abdalla2005simple} and the SPAKE2 library, we also enforce identities---again as byte strings---at the level of \ac{pake} instances, which can refer to a username, user ID or server names, to name a few. As detailed in \Cref{sec:instantiation-based-on-spake2}, the public key fingerprints could be included in the transcript and thus in the input of the hash function computing the intermediate shared key before \ac{kc}, however the SPAKE2 \ac{api} accepts only the user IDs and the weak password. We deal with this using the \ac{kc} step and the inclusion of the public key fingerprints as associated data into the HMAC-authenticated message.

For further information on the details of the underlying SPAKE2 implementation, we refer the reader to the corresponding documentation by Warner \cite{spake2python}.

\subsection{PAKE protocol carried out over email}

We have implemented the email-based approach suggested in \Cref{sec:transport-mechanism}, mainly because it corresponds to the solution that preservers compatibility and interoperability without imposing any additional requirements on standard email exchange solutions. 
\ac{pakemail} essentially makes use of email messages and attachments as transport mechanism 
for exchanging cryptographic messages and key confirmation tags belonging to \ac{pake} protocol sessions as well as other data such as public keys that are to be authenticated by \ac{pake} messages, effectively implementing the communication channel via mailboxes. In the case of secure messaging, the networking would be rather trivial given that most current solutions make use of intermediary servers, which in our case can be untrusted.

\subsection{Implemented scenarios}

The solution provides
\ac{pake} clients and email services designed to deal with the requirements of \ac{pake} exchanges and state maintenance in a decentralized and distributed computing setting. The \ac{pake} clients have been implemented such that they take on either the role of an  ``initiator'' or that of a ``responder'', consistent with the original SPAKE2 protocol design and the requirements of the SPAKE2 Python \ac{api}.

Moreover, we provide a module containing easy to use executable implementations of the following scenarios: $(i)$ a local execution of two independent threads of \ac{pake} clients running a \ac{pake} session with key confirmation, followed by some cryptographic tasks using the established key; $(ii)$ an online execution of two clients (an initiator and a responder instance) running on the same hardware but routing their messages via email exchanges and attachments, currently implemented to work with Gmail but adapting it to other services would simply amount to providing the appropriate access data, e.g., the corresponding mail server credentials and port numbers; $(iii)$ and $(iv)$ provide the execution of initiator and responder instances, respectively, on two different machines, again using email as transport mechanism.

\subsection{Performance}

In terms of performance, the main scenario of interest, namely that of running two separate instances of \ac{pakemail} on two different machines, carrying out a \ac{pake} protocol with explicit key confirmation over Gmail, requires $\approx 3 \cdot 10^{-3}$ seconds, averaged over 10 runs. The results were obtained from executions on two laptops running at 1.6 GHz (Dual-Core Intel Core i5) with 8 GB of RAM, 256 KB and 4 MB of L2 and L3 cache, respectively.

Given the setting for which this approach is designed, i.e., distributed peer-to-peer connections between entities running point-to-point \ac{pake} sessions, we consider the current overall execution time to be fast enough for all practical purposes. \Cref{tab:exec-time} provides a concise comparison of execution times for pure SPAKE2 sessions with its \ac{pakemail} counterpart, providing some information on the overall overhead added by our email-based networking and other non-\ac{pake} computations.

\begin{table}[th]
\centering
\caption{Execution time comparison averaged over 10 runs}
\label{tab:exec-time}
\resizebox{0.8\textwidth}{!}{%
\begin{tabular}{c|c|c|c}
\hline
\textbf{Group} & \textbf{Pure SPAKE2} & \textbf{Local \ac{pakemail}} & \textbf{\ac{pakemail} via Gmail} \\ \hline
Curve25519                    & 26 ms                & 50 ms                        & 350 ms                    \\
\end{tabular}%
}
\end{table}

Note that once both parties have entered their passwords, the added networking overhead due to email exchanges triggered by PakeMail will arguably not be perceptible by users given the inherent delay in email exchanges.

Finally, in terms of the underlying SPAKE2 library's performance on the same hardware, the average execution times using Curve25519 and 1024/2048/3072-bit integer groups are 26 ms, 9 ms, 42.1 ms and 82.6 ms, respectively. The delta would simply contribute additively to the PakeMail executions as the additional overhead incurred by switching to different representations is independent from the details of PakeMail.

\subsection{Further design and security considerations}

Due to the nature of the current proof of concept implementation, certain design decisions have been made simply to ensure the implementation of a functional tool capable of demonstrating the feasibility, usability and efficiency of the proposed approach. However, a mature and robust implementation would have to account for a number of nuances. For instance, for the purpose of our proof of concept, we simply use \ac{uuid} numbers along with other user identifiers, which are stored in the email subject, to synchronize and map initiator and responder messages belonging to the same session to one another, coupled with a persistent per client session history to track and resolve sessions. A robust networking component capable of addressing distributed systems corner cases such as deadlocks and race conditions remains to be done.

Regarding the cryptographic details of the implementation, it should be pointed out that a secure and scalable industrial implementation would have to at the very least rely on a constant-time implementation of the \ac{pake} library as the currently used SPAKE2 library is by no means constant-time and is thus vulnerable to timing attacks.

Finally, note that dedicated optimization efforts remain to be done as future work. Clearly, the alternative transport mechanism based on intermediary servers, enabling more natural communication channels and networking, would lead to far lower communication overhead, albeit at the cost of somewhat hampering interoperability and compatibility, unless projects such as Matrix\footnote{\url{https://matrix.org/docs/spec/}} and MLS\footnote{\url{https://messaginglayersecurity.rocks/}} gain widespread adoption.

\section{Security and Low-Entropy Secrets}\label{sec:security}

The schemes considered thus far come with proofs of security, see \Cref{tab:pake} for the corresponding models and assumptions. The security guarantees can be traced back to the core properties of \ac{pake}s: they can in effect fulfill the role of \ac{zk} proof of knowledge schemes such that a run of the protocol does not leak any information on the password and upon termination only reveals whether the secrets were equal; they resist offline dictionary attacks against passive and active adversaries, and online guessing attacks by limiting adversarial tests to one password per run; compromised session keys do not compromise the security of other established session keys; depending on the choice of \ac{pake}, \ac{fs} would ensure that session keys remain secure in case of password disclosure.

The only way for \adver to gain knowledge about the secret would be via active online guessing attempts, typically dealt with by fixing a limit on the number of failed attempts, e.g., \ac{smp} in \ac{otr}. As we previously discussed, the possibility of making \ac{pake}s auditable can be used to mitigate this class of attacks by distinguishing between failed adversarial attempts and network failures to minimize the adversary's tries to one, under the assumption of correct input entry by honest users.

\paragraph{\textbf{Low-entropy secret agreement.}}\label{sec:inband-secret}
Our proposal does come with a caveat, namely the need for either presharing or agreeing on a low-entropy secret in-band. As already discussed in \cite{alexander2007improved}, users can either share a secret over a secure channel, e.g. \ac{oob}, or agree on one via an in-band solution without revealing sensitive information about the secret itself, e.g., \A asking \B to use the name of their favorite restaurant. The user interface of a tool implementing this could warn users not to include the secret itself, similar to standard email warnings reminding users to attach documents in case they have mentioned it in the body of the message.

Assuming already bootstrapped authentication to avoid circularity, another possibility would be to use another already authenticated and secure channel to agree on a secret.
For instance, given the widespread use of tools such as Signal, parties could simply use it to agree on a secret for a one-time entity authentication of their secure email solution. While it may not be appealing from a theoretical point of view, due to the assumption of there being an already authenticated and secure channel, practically speaking, this approach would in fact provide a realistic and usable solution. 

\paragraph{\textbf{Usability aspects.}}
Particular attention must be paid to the implementation of an adequate interface for entering the low-entropy secret, along with the corresponding documentation and manuals with simple explanations for users.
A lesson learned from a usability study on the \ac{otr}/\ac{smp} tool \cite{usabilityOTR} stresses the need for further research on how to guide users towards establishing a secure shared human-memorable secret. 

For instance, adding a list pre-populated with questions might serve to reduce user effort by allowing them to choose one from the list, or as a guide for users to generate similar questions. The questions should not lead to evident answers or to answers belonging to very small known sets, such as ``yes/no'' or colors, as such cases increase the successful guessing probability of the adversary. Another measure for dealing with disparities due to letter cases would be to for example simply convert the secret to upper-case, at the cost of reducing entropy.

\section{Further directions}
\label{sec:conclusions}

A clear and promising line of future work consists of improving the current implementation and adding the various enhancements discussed here.

Producing secure implementations of cryptographic primitives and protocols is a notoriously difficult task. Consequently, over the past decades, a considerable amount of research in formal verification has focused on developing techniques for ensuring that 
security software preserves the security guarantees of the underlying cryptographic constructions. 
Although our solution builds on provably secure cryptographic constructions, the actual implementation makes use of cryptographic software that has not been proven secure.
Therefore, pursuing the development of a verified implementation of a \ac{pake} protocol would be another promising research direction. This could be achieved using dedicated languages such as F* \cite{fstar}, which has been used, among other things, to produce a verified reference implementation of the TLS (1.2) protocol \cite{bhargavan2013implementing}.

Alternatively, a robust \ac{pake} implementation in a language designed for performance and safety such as RUST would be yet another viable path. An initial rough implementation of SPAKE2 in RUST is already available and subject to ongoing work\footnote{\url{https://github.com/RustCrypto/PAKEs}}.

Follow-up theoretical work on all the suggested cryptographic enhancements and implementations thereof represents another line of research. In particular, given the fact that mature \ac{pake} implementations are quite rare, we consider further theoretical work on the design and analysis of a quantum-secure \ac{pake}, proven secure in the QROM, accompanied by an actual implementation to be worth pursuing. Similarly, to the best of our knowledge, an implementation, let alone practical and efficient, of the secure secret storage and retrieval tasks (e.g., using \ac{ppss} or OPAQUE) represents yet another promising line of work.

Moreover, research on effective and usable methods for assisting users in agreeing on low-entropy secrets while reducing the mental effort and the likelihood of mistakes, is also encouraged. 
Other interesting directions include the application of \ac{pake} to authentication for encrypted mailing lists, and studying the possibility of sharing/synchronizing existing trust assignments for contacts across different services---e.g., from Signal to \pep or vice versa. In this case, once an entity is trusted in one application, other applications that recognize this entity could inherit the trust stored in the user's device; clearly, it is vital to do this in a secure and privacy-preserving manner.

\bibliographystyle{splncs04}
\bibliography{bibliography}

\end{document}